\documentclass[aps,prc,amsfonts,preprintnumbers,superscriptaddress,showpacs,nofootinbib]{revtex4}

\usepackage{graphics}
\usepackage{psfrag}
\usepackage{epsfig}
\usepackage{epsf}
\usepackage{float}

\newcommand{\fet}[1]{\mbox{\boldmath $#1$}}

\newcommand{\beq}{\begin{equation}}
\newcommand{\eeq}{\end{equation}}
\newcommand{\beqa}{\begin{eqnarray}}
\newcommand{\eeqa}{\end{eqnarray}}
\newcommand{\nn}{\nonumber \\ }

\begin{document}

{\tiny
\preprint{FZJ-IKP-TH-2007-36}
\preprint{HISKP-TH-07/29}
}

\title{Isospin-breaking two-nucleon force with explicit $\Delta$-excitations}

\author{E.~Epelbaum}
\email[]{Email: e.epelbaum@fz-juelich.de}
\affiliation{Forschungszentrum J\"ulich, Institut f\"ur Kernphysik 
(Theorie), D-52425 J\"ulich, Germany}
\affiliation{Universit\"at Bonn, Helmholtz-Institut f{\"u}r
  Strahlen- und Kernphysik (Theorie), D-53115 Bonn, Germany}
\author{H.~Krebs}
\email[]{Email: hkrebs@itkp.uni-bonn.de}
\affiliation{Universit\"at Bonn, Helmholtz-Institut f{\"u}r
  Strahlen- und Kernphysik (Theorie), D-53115 Bonn, Germany}
\author{Ulf-G.~Mei{\ss}ner}
\email[]{Email: meissner@itkp.uni-bonn.de}
\homepage[]{URL: www.itkp.uni-bonn.de/~meissner/}
\affiliation{Universit\"at Bonn, Helmholtz-Institut f{\"u}r
  Strahlen- und Kernphysik (Theorie), D-53115 Bonn, Germany}
\affiliation{Forschungszentrum J\"ulich, Institut f\"ur Kernphysik 
(Theorie), D-52425 J\"ulich, Germany}
\date{\today}

\begin{abstract}
We study the leading isospin-breaking contributions to the two-nucleon
two-pion exchange potential due to explicit $\Delta$ degrees of freedom in  chiral
effective field theory. In particular, we find important contributions due
to the delta mass splittings to the charge symmetry breaking potential 
that act opposite to the effects induced by the
nucleon mass splitting.
\end{abstract}

\pacs{13.75.Cs,21.30.-x}

\maketitle

\vspace{-0.2cm}

%%%%%%%%%%%%%%%%%%%%%%%%%%%%%%%%%%%%%%%%%%%%%%%%%%%%%%%%%%%%%%%%%%%%%%%%%%%%%%%%%
\section{Introduction}
\def\theequation{\arabic{section}.\arabic{equation}}
\setcounter{equation}{0}
\label{sec:intro}

Isospin-violating (IV) two- (2NF) and three-nucleon forces (3NF) have attracted  a lot of
interest in the 
recent years in the context of chiral effective field theory (EFT), see
\cite{Epelbaum:2005pn} and references therein. In the two-nucleon sector, IV
one-pion \cite{VanKolck:1993ee,vanKolck:1996rm,Walzl:2000cx,Friar:2004ca,Epelbaum:2005fd},  
two-pion \cite{Friar:1999zr,Walzl:2000cx,Niskanen:2001aj,Friar:2003yv,Epelbaum:2005fd},
one-pion-photon \cite{vanKolck:1997fu,Kaiser:2006ws} and two-pion-photon
exchange \cite{Kaiser:2006ws,Kaiser:2006ck,Kaiser:2006na}
potentials as well as short-range contact interactions \cite{Walzl:2000cx,Epelbaum:2005fd}
have been studied in this framework up to rather 
high orders in the EFT expansion, as also reviewed in \cite{Epelbaum:2005pn}.
In addition, the leading and subleading IV 3NF have been worked
out in Refs.~\cite{Friar:2004rg,Epelbaum:2004xf}. As found in
Ref.~\cite{Epelbaum:2005fd}, the charge-symmetry breaking (CSB) two-pion exchange
potential (TPEP) in EFT without explicit $\Delta$ degrees of freedom exhibits an
unnatural convergence pattern with the (formally) subleading contribution
yielding numerically the dominant effect. This situation is very similar to the 
isospin-conserving TPEP, where the unnaturally strong contribution at
next-to-next-to-leading order in the chiral expansion can be attributed to the
large values of the low-energy constants (LECs) $c_{3.4}$  accompanying the
subleading $\pi \pi NN$ vertices. The large values of these LECs are well
understood in terms of resonance saturation \cite{Bernard:1996gq}. In particular, 
the $\Delta$-isobar provides the dominant (significant)
contribution to $c_3$ ($c_4$). Given its low excitation energy,
$\Delta  \equiv m_\Delta - m_N = 293$ MeV, and strong 
coupling to the $\pi N$ system, one expects that the explicit inclusion of 
the $\Delta$
in EFT utilizing e.g.~the so-called small scale expansion (SSE) 
\cite{Hemmert:1997ye}. Such a scheme allows to resum a certain class of 
important contributions and 
improves the convergence as compared to the delta-less theory. For the
isospin-invariant TPEP, the improved convergence in the delta-full theory has
indeed been verified via explicit calculations
\cite{Kaiser:1998wa,Krebs:2007rh}. It is natural to expect that
including the $\Delta$-isobar as an explicit degree of freedom will also improve
the convergence for the IV TPEP. In our recent work \cite{Epelbaum:2007sq} we
already made an important step in this direction and analyzed the delta
quartet mass splittings in chiral EFT. In addition, we worked out the leading
$\Delta$-contribution to IV 3NF. As expected based on resonance saturation, a
significant part of the (subleading) CSB 3NF proportional to LECs
$c_{3,4}$ in the delta-less theory was demonstrated to be shifted to the
leading order in the delta-full theory. We also found important effects which go beyond
resonance saturation of LECs $c_{3,4}$.    

In this paper we derive the leading $\Delta$-contributions to the IV two-pion
exchange 2NF and compare the results with the calculations based on the
delta-less theory. Our manuscript is organized as follows: in sec.~\ref{sec1}, we
briefly discuss our formalism and present expressions for the IV TPEP in
momentum space. We Fourier transform the potential into coordinate space and
compare the obtained results at leading order in the delta-full theory with
the ones found at subleading order in the delta-less theory
\cite{Epelbaum:2005fd} in sec.~\ref{sec2}. 
Our work is summarized in sec.~\ref{sec3}.

%%%%%%%%%%%%%%%%%%%%%%%%%%%%%%%%%%%%%%%%%%%%%%%%%%%%%%%%%%%%%%%%%%%%%%%%%%%%%%%%%
\section{$\Delta$-contributions to the leading isospin-breaking two-pion
  exchange potential}
\def\theequation{\arabic{section}.\arabic{equation}}
\setcounter{equation}{0}
\label{sec1}

\begin{figure}[tb]
\vskip 1 true cm
%  \begin{center} 
%    \epsfxsize=4.3cm
%    \epsffile{fig2.eps}
\includegraphics[width=13.0cm,keepaspectratio,angle=0,clip]{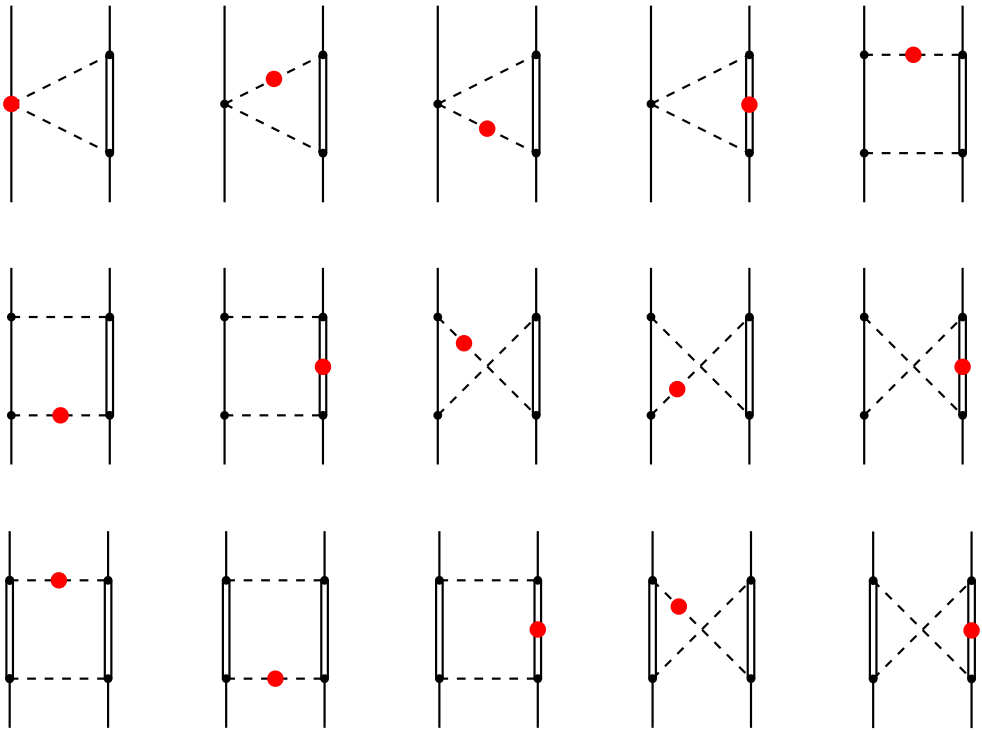}
    \caption{
         Leading isospin-breaking contributions to the $2\pi$-exchange NN
         potential due to intermediate $\Delta$-excitation. Solid, dashed
         and double lines represent nucleons, pions and deltas,
         respectively. Solid dots and filled 
         circles refer to  the leading isospin-invariant and isospin-breaking
         vertices, in order. Diagrams 
         resulting by the interchange of the nucleon lines are not shown. 
\label{fig1} 
 }
%  \end{center}
\end{figure}

To obtain the leading isospin-violating $\Delta$-contributions to the TPEP one
has to evaluate the triangle, box and crossed-box diagrams with one insertion
of isospin-breaking pion, nucleon and delta mass shifts $\delta M_\pi$, $\delta
m_N$, $\delta m_\Delta^1$ and $\delta m_\Delta^2$, respectively.  
In addition, one needs to consider 
triangle diagrams involving the leading strong IV $\pi \pi NN$ vertex. Following
Ref.~\cite{Epelbaum:2005pn}, we also include the leading  electromagnetic
$\pi \pi NN$ vertices. Notice that these IV $\pi \pi NN$ vertices do not contribute to
the 3NF with the intermediate $\Delta$-excitation and were not
considered in \cite{Epelbaum:2007sq}. We follow the
same strategy as in Refs.~\cite{Epelbaum:2007sq,Friar:2004ca} and eliminate
the neutron-proton 
mass difference term from the effective Lagrangian in favor of new
isospin-violating vertices proportional to $\delta m_N$. This allows one to
directly use the Feynman graph technique to derive the
corresponding NN potential and thus considerably simplifies the calculations. 
The leading $\Delta$-contribution to the IV TPEP arises from Feynman diagrams
shown in Fig.~\ref{fig1}. 
The Feynman rules for the relevant isospin-invariant vertices can be
found e.g. in Ref.~\cite{Bernard:1995dp}, see also \cite{Krebs:2007rh}. 
To the order we are working, the Feynman rules for IV vertces after
eliminating the neutron-proton mass shift have the following form.
\begin{itemize}
\item
Two pions, no nucleons, no $\Delta$:
\beq
i \delta M_\pi^2 \, \delta_{a3} \delta_{b3}  + \delta m_N \epsilon_{3ab} \,
(q_2 - q_1 ) \cdot v \,.
\eeq
Here, $v_\mu$ is the baryon four-velocity, $q_1$ and $q_2$ denote the incoming
pion momenta with the isospin quantum numbers $a$ and $b$, respectively,
$\delta M_\pi^2 \equiv M_{\pi^\pm}^2 - M_{\pi^0}^2 $ is the difference of the
squared charged and neutral pion masses,
and $\delta m_N \equiv m_p - m_n$ is the neutron-proton mass
difference. Here and in what follows, we express, whenever possible, the LECs accompanying
isospin-violating vertices in terms of the pion, nucleon and delta mass
shifts.    
\item
No pions, no nucleons, one $\Delta$:
\beq
i \, \Big[  \left( \delta m_\Delta^1 - 3 \delta m_N \right) \frac{\tau^3}{2} \delta_{ij}
+ \delta m_\Delta^2 \frac{3}{4} \delta_{i3} \delta_{j3} \Big] g_{\mu \nu}\,,
\eeq
where $i, \; j$ and $\mu , \; \nu$ refer to the isospin and Lorentz indices of
the Rarita-Schwinger field and $\tau^i$ denotes the Pauli isospin matrix with
the isospin index $i$. Further, $\delta m_\Delta^1$ and $\delta m_\Delta^2$
are the two isospin-violating delta mass shifts introduced and analyzed in
Ref.~\cite{Epelbaum:2007sq}.    
\item
Two pions, one nucleon, no $\Delta$:
\beq
i\,\frac{\delta m_N^{\rm str}}{2 F_\pi^2} \left(\tau^a \delta_{b3} + \tau^b \delta_{a3} -
  \tau^3 \delta_{ab}
\right) + i \, 2 e^2 f_1 \, \delta_{a3} \delta_{b3}\,,
\eeq
where $F_\pi$ is the pion decay constant and $f_1$ is a
LEC accompanying one of the leading-order electromagnetic
operators, see Ref.~\cite{Meissner:1997ii,Gasser:2002am} for more details. 
Notice that the last term in the above expression leads to the IV 3NF, see
\cite{Epelbaum:2004xf}, but does not generate a IV two-pion exchange 2NF at the order
considered here, see also \cite{Epelbaum:2005fd} for
the same conclusion made using the delta-less EFT.   
\end{itemize}

The leading IV NN potential in
the center-of-mass system (CMS) can be conveniently written in the form:
\beq
\label{2PEdec}
V = \tau_1^3 \tau_2^3 \; \Big[ 
V_C^{\rm II} + V_S^{\rm II } \, \vec \sigma_1 \cdot \vec \sigma_2 
+  V_T^{\rm II } \, \vec \sigma_1 \cdot \vec q \,\, \vec \sigma_2 \cdot \vec q
\Big] 
+  ( \tau_1^3 +  \tau_2^3 ) \; \Big[ 
V_C^{\rm III} + V_S^{\rm III} \, \vec \sigma_1 \cdot \vec \sigma_2 
+  V_T^{\rm III} \, \vec \sigma_1 \cdot \vec q \,\, \vec \sigma_2 \cdot \vec q
\Big] \,,
\nonumber
\eeq
where  $\vec p$ and $\vec p
\, '$  are the initial and final CMS momenta, $\vec \sigma_i$ ($\fet \tau_i$)
refers to the spin (isospin) matrices of  nucleon~$i$ and  
$\vec q \equiv \vec p \, ' - \vec p$,
$\vec k \equiv \frac{1}{2} (\vec p \, ' + \vec p \, )$. 
Further, the superscripts $C$, $S$, $T$ of the scalar functions 
$V_C$, $V_{S}$ and $V_T$ denote the central, spin-spin and tensor 
components while the superscripts refer to the class-II and class-III
isospin-violating 2NFs in the notation of Ref.~\cite{Henley:1979aa}. The
class-II interactions conserve charge symmetry and are often referred to as
charge-independence breaking while the class-II 2NF are charge-symmetry breaking,
see also Ref.~\cite{Epelbaum:2004xf}. Notice that at this order there are no class-IV
contributions to the TPEP which lead to isospin-mixing. Evaluating the
diagrams shown in Fig.~\ref{fig1} and utilizing the spectral function
regularization framework \cite{Epelbaum:2003gr} we obtain the following
expressions for the scalar 
functions in Eq.~(\ref{2PEdec}): 
\begin{itemize}
\item $\Delta$-excitation in the triangle graphs:
\beqa
\label{triangle}
V_C^{\rm II}&=&-\frac{ h_A^2 }{144 F_\pi^4 \pi ^2 \Delta  \Sigma } \left(2 \Delta
  ^2 \Sigma  (4 \Delta  (3 \Delta  \delta 
  m_\Delta^2+\delta M_\pi^2)-3 \delta m_\Delta^2 \Sigma
  ) D^{\tilde \Lambda}(q)+2 \left(-M_\pi^2+\Delta ^2\right) \left(2 \Delta  (9 \Delta
    \delta m_\Delta^2+
\delta M_\pi^2)\right.\right.\nn &&
\left.\left. -3 \delta m_\Delta^2 \omega ^2\right) H^{\tilde \Lambda}(q)+6 \Sigma  \left(8
    \Delta ^2 \delta m_\Delta^2+\Delta  \delta M_\pi^2+\delta m_\Delta^2
    \left(\Sigma -\omega 
^2\right)\right) L^{\tilde \Lambda}(q)\right) \,, \nn
V_C^{\rm III}&=&
-\frac{h_A^2 \Delta }{216 F_\pi^4 \pi ^2}
\left(6 \Delta ^2 (7 \delta m_N-2 \delta m_N^{\rm str}-5 \delta m_\Delta^1)+\right.
(-15 \delta m_N+6 \delta m_N^{\rm str}+5 \delta m_\Delta^1) 
\left.\left.\left(2 \Delta ^2+\Sigma \right)\right) D^{\tilde \Lambda}(q)\right.\nn &&\left.-
\frac{h_A^2}{216 F_\pi^4 \pi ^2 \Delta  \Sigma }
\left(-M_\pi^2+\Delta ^2\right) \right.
\left(6 \Delta ^2 (11 \delta m_N-4 \delta m_N^{\rm str}-5 \delta m_\Delta^1)+\right.
\left.(-15 \delta m_N+6 \delta m_N^{\rm str}+5 \delta m_\Delta^1) \omega ^2\right) 
H^{\tilde \Lambda}(q)\nn &&-
\frac{h_A^2}{108 F_\pi^4 \pi ^2 \Delta }
\left(3 \Delta ^2 (9 \delta m_N-3 \delta m_N^{\rm str}-5 \delta m_\Delta^1)+\right.
\left.M_\pi^2 (-15 \delta m_N+6 \delta m_N^{\rm str}+5 \delta m_\Delta^1)\right) 
L^{\tilde \Lambda}(q),\nn
V_S^{\rm II}&=&V_T^{\rm II}=V_S^{\rm III}=V_T^{\rm III}=0
\eeqa
\item Single $\Delta$-excitation in the box and crossed-box graphs:
\beqa
\label{box_single}
V_C^{\rm II}&=&\frac{g_A^2 h_A^2}{144 F_\pi^4 \pi ^2 \Delta ^3 \Sigma  \omega ^2}
\left(-\Delta ^2 \Sigma ^2 (8 \Delta  (3 \Delta  \delta m_\Delta^2+\delta
  M_\pi^2)+3 \delta m_\Delta^2 \Sigma 
) \omega ^2 D^{\tilde \Lambda}(q)\right.\nn &&-
\left(M_\pi^2-\Delta ^2\right) \omega ^2 \left(72 \Delta ^4 \delta m_\Delta^2+40
  \Delta ^3 \delta M_\pi^2-3 \delta m_\Delta^2 \omega
  ^4-6 \Delta ^2 \delta m_\Delta^2 \left(\Sigma +\omega ^2\right)-2 \Delta  
\delta M_\pi^2 \left(\Sigma
+5 \omega ^2\right)\right) H^{\tilde \Lambda}(q)\nn &&+
\Sigma  \left(2 \Delta  \delta M_\pi^2 \left(2 \Delta ^2+\Sigma \right)^2+2
  \Delta  \left(42 \Delta ^3 \delta m_\Delta^2+23 
\Delta ^2 \delta M_\pi^2+6 \Delta  \delta m_\Delta^2 \Sigma +4 \delta M_\pi^2
\Sigma \right) \omega ^2\right.\nn &&+ 
\left.\left.(-2 \Delta  (3 \Delta  \delta m_\Delta^2+5 \delta M_\pi^2)+3
    \delta m_\Delta^2 \Sigma ) \omega ^4-3 
\delta m_\Delta^2 \omega ^6\right) L^{\tilde \Lambda} (q)\right) \,, \nn 
V_S^{\rm II}&=& - q^2 V_T^{\rm II} = \frac{g_A^2 h_A^2 }{576 F_\pi^4 \pi  \Delta
  ^2}  q^2 \left(8 \Delta  \delta M_\pi^2+3 \delta 
    m_\Delta^2 \omega ^2\right) A^{\tilde \Lambda}(q) \,, \nn
V_C^{\rm III}&=&-\frac{g_A^2 h_A^2}{864 F_\pi^4 \pi ^2 \Delta ^3 \Sigma  \omega ^2} 
 \left(10 \pi  \Delta  (3 \delta m_N-\delta m_\Delta^1) \Sigma  \left(2
    \Delta ^2+\Sigma \right)^2 
\omega ^2 A^{\tilde \Lambda}(q)+2 \Delta ^2 \Sigma ^2 \left(8 \Delta ^2 (3 \delta m_N-5
  \delta m_\Delta^1)\right.\right.\nn &&
\left.\left.+5 (3 \delta m_N-\delta m_\Delta^1) \Sigma \right) \omega ^2 D^{\tilde \Lambda}(q)+\right.
2 \left(-M_\pi^2+\Delta ^2\right) \omega ^2 \left(120 \Delta ^4 (\delta m_N+\delta
  m_\Delta^1)+5 (3 \delta m_N- 
\delta m_\Delta^1) \omega ^4\right.\nn && 
\left.+2 \Delta ^2 \left(3 \delta m_N \left(\Sigma -15 \omega ^2\right)-5
    \delta m_\Delta^1 \left(\Sigma +\omega 
^2\right)\right)\right) H^{\tilde \Lambda}(q)+2 \Sigma  \left(24 \Delta ^2 \delta m_N \left(2
\Delta ^2+\Sigma 
  \right)^2\right.\nn &&
\left.+4 \Delta ^2 \left(\Delta ^2 (33 \delta m_N+
35 \delta m_\Delta^1)+(9 \delta m_N+5 \delta m_\Delta^1) \Sigma \right) \omega
^2-\left.5 \left(2 \Delta ^2 (9 \delta m_N+\delta
    m_\Delta^1)\right.\right.\right.\nn && 
\left.\left.\left.+(3 \delta m_N-\delta m_\Delta^1) \Sigma
\right) \omega ^4+5 (3 \delta m_N-\delta m_\Delta^1) \omega
^6\right) L^{\tilde \Lambda} (q)\right)\,, \nn
V_S^{\rm III}&=&- q^2 V_T^{\rm III} = \frac{g_A^2 h_A^2}{3456 F_\pi^4 \pi ^2
  \Delta ^3 \Sigma } q^2 
\left(10 \pi  \Delta  (3 \delta m_N-\delta m_\Delta^1) 
  \Sigma  \omega ^2 A^{\tilde \Lambda}(q)+2 \Delta ^2 \Sigma  \left(4 \Delta ^2 
(3 \delta m_N-5 \delta m_\Delta^1)\right.\right.\\ &&
\left.\left.+5 (3 \delta m_N-\delta m_\Delta^1) \omega ^2\right) D^{\tilde \Lambda}(q)-5 (3
  \delta m_N-\delta m_\Delta^1) \left(2 \left(-M_\pi^2+\Delta ^2\right) \left(4
      \Delta ^2-\omega ^2\right) H^{\tilde \Lambda}(q)-4 M_\pi^2 
\Sigma  L^{\tilde \Lambda} (q)\right)\right)\,.  \nonumber
\eeqa
\item Double $\Delta$-excitation in the box and crossed-box graphs:
\beqa
\label{box_double}
V_C^{\rm II}&=&-\frac{h_A^4}{1296 F_\pi^4 \pi ^2 \Delta ^3 \Sigma  \left(4 \Delta
    ^2-\omega ^2\right)}
\left(-16 \Delta ^3 \left(-M_\pi^2+\Delta ^2\right) (3 \Delta  \delta
  m_\Delta^2+\delta M_\pi^2) \Sigma \right.\nn && \left. +
2 \Delta ^2 \Sigma
 \left(32 \Delta ^3 (3 \Delta  \delta m_\Delta^2+\delta M_\pi^2)-8
   \Delta  (6 \Delta  \delta m_\Delta^2+
\delta M_\pi^2) \Sigma -3 \delta m_\Delta^2 \Sigma ^2\right) \right.\nn &&
\times\left(4 \Delta ^2-\omega ^2\right) D^{\tilde \Lambda}(q)+\left(8 \Delta ^3 \left(8 \Delta
    ^4 (69 \Delta  \delta m_\Delta^2+17 
    \delta M_\pi^2)+2 \Delta ^2 (63 \Delta  
\delta m_\Delta^2+17 \delta M_\pi^2) \Sigma +\delta M_\pi^2 \Sigma
^2\right)\right.\nn &&- 
2 \Delta  \left(16 \Delta ^4 (87 \Delta  \delta m_\Delta^2+22 \delta
  M_\pi^2)+2 \Delta ^2 (87 \Delta   
\delta m_\Delta^2+25 \delta M_\pi^2) \Sigma -(6 \Delta  \delta
m_\Delta^2+\delta M_\pi^2) \Sigma ^2\right) \omega ^2\nn &&+ 
\left.4 \Delta  \left(129 \Delta ^3 \delta m_\Delta^2+37 \Delta ^2 \delta
    M_\pi^2+3 \Delta  \delta m_\Delta^2 
\Sigma +2 \delta M_\pi^2 \Sigma \right) \omega ^4+(-2 \Delta  (6 \Delta
\delta m_\Delta^2+5 \delta M_\pi^2)+3 
\delta m_\Delta^2 \Sigma ) \omega ^6\right.\nn &&\left. -3 \delta m_\Delta^2 
\omega ^8\right) H^{\tilde \Lambda}(q)+
2 \Sigma  \left(1296 \Delta ^6 \delta m_\Delta^2+320 \Delta ^5 \delta
  M_\pi^2+12 \Delta ^3 \delta M_\pi^2 
\left(3 \Sigma -10 \omega ^2\right)+156 \Delta ^4 \delta m_\Delta^2
\left(\Sigma -3 \omega ^2\right)\right.\nn &&\left. +3 \delta m_\Delta^2 \omega
^4 \left(-\Sigma +\omega ^2\right)-\left.2 \Delta  \delta M_\pi^2 \left(\Sigma
    ^2+4 \Sigma 
      \omega ^2-5 \omega ^4\right)-6 \Delta ^2 
\delta m_\Delta^2 \left(\Sigma ^2+4 \Sigma  \omega ^2-4 \omega
  ^4\right)\right) L^{\tilde \Lambda} (q)\right)  \,, \nn
V_S^{\rm II}&=&- q^2 V_T^{\rm II} = -\frac{h_A^4}{10368 F_\pi^4 \pi ^2 \Delta ^3
  \Sigma }q^2 \left(2 \Delta ^2 \Sigma  \left(12 \Delta ^2 \delta
    m_\Delta^2-8 
    \Delta  \delta M_\pi^2-3 
\delta m_\Delta^2 \omega ^2\right) D^{\tilde \Lambda}(q)\right.\nn &&+
\left.\left(4 \Delta ^2 (-4 \Delta  \delta M_\pi^2+3 \delta m_\Delta^2 \Sigma
    )-3 \delta m_\Delta^2 \left(8 
\Delta ^2+\Sigma \right) \omega ^2+3 \delta m_\Delta^2 \omega 
^4\right) H^{\tilde \Lambda}(q)-12 M_\pi^2 \delta m_\Delta^2 \Sigma  L^{\tilde \Lambda} (q)\right)\,, \nn
V_C^{\rm III}&=&\frac{h_A^4}{3888 F_\pi^4 \pi ^2 \Delta ^3 \Sigma  \left(4 \Delta
    ^2-\omega ^2\right)} 
\left(48 \Delta ^4 \left(-M_\pi^2+\Delta ^2\right) (7 \delta m_N-5 \delta
  m_\Delta^1) \Sigma \right.\nn &&\left. -2 \Delta ^2 \Sigma  \left(96 \Delta
^4 (7 \delta m_N-5 \delta m_\Delta^1)+16 \Delta ^2 (-3
\delta m_N+5 \delta m_\Delta^1) \Sigma +
5 (3 \delta m_N-\delta m_\Delta^1) \Sigma ^2\right) \right.\nn && 
\times\left(4 \Delta ^2-\omega ^2\right) D^{\tilde \Lambda}(q)+\left(-192 \Delta ^8 (89 \delta
  m_N-75 \delta m_\Delta^1)+5 (3 \delta m_N-\delta m_\Delta^1) \omega ^6
  \left(\Sigma 
-\omega ^2\right)\right.\nn &&\left. -16 \Delta ^6 \left(297 \delta m_N \Sigma
-235 \delta m_\Delta^1 \Sigma -714 \delta m_N \omega ^2+590 
\delta m_\Delta^1 \omega ^2\right)+\right. 4 \Delta ^4 \left(8 (-9 \delta
m_N+5 \delta m_\Delta^1) \Sigma ^2\right.\nn &&\left. +5 (93 \delta m_N-71
\delta m_\Delta^1) 
\Sigma  \omega ^2+3 (-229 \delta m_N+175 \delta m_\Delta^1) \omega
^4\right)+\left.4 \Delta ^2 \omega ^2 \left(-3 \delta m_N \left(\Sigma ^2+19
    \Sigma  \omega ^2-25 \omega ^4\right)\right.\right.\nn &&\left.\left. +5
  \delta m_\Delta^1 
\left(\Sigma ^2+7 \Sigma  \omega ^2-9 \omega ^4\right)\right)\right) H^{\tilde \Lambda}(q)-2
\Sigma  \left(80 \Delta ^6 (63 \delta m_N-53 \delta m_\Delta^1)+5 (3 \delta
  m_N-\delta m_\Delta^1) \omega 
^4 \left(\Sigma -\omega ^2\right)\right.\nn &&\left. +4 \Delta ^4 \left(129
  \delta m_N \Sigma -115 \delta m_\Delta^1 \Sigma -495 \delta m_N 
\omega ^2+405 \delta m_\Delta^1 \omega ^2\right)+\left.2 \Delta ^2 \left(3 (-7 
  \delta m_N+5 \delta m_\Delta^1) \Sigma ^2\right.\right.\right.\nn
&&\left.\left.\left. + 
12 (-7 \delta m_N+ 5 \delta m_\Delta^1) \Sigma  \omega ^2+40 (3 \delta m_N-2
\delta m_\Delta^1) 
\omega ^4\right)\right) L^{\tilde \Lambda} (q)\right) \,,  \nn
V_S^{\rm III}&=&-q^2 V_T^{\rm III} = -\frac{h_A^4 }{31104 F_\pi^4 \pi ^2 \Delta ^3
  \Sigma }q^2 \left(2 
  \Delta ^2 \Sigma  \left(4 \Delta ^2 (57 \delta m_N-35 \delta m_\Delta^1)+5
    (3 \delta m_N- 
\delta m_\Delta^1) \omega ^2\right) D^{\tilde \Lambda}(q)\right.\nn &&+
\left(64 \Delta ^4 (-9 \delta m_N+5 \delta m_\Delta^1)-20 \Delta ^2 (3 \delta
  m_N-\delta m_\Delta^1) \left(\Sigma 
-2 \omega ^2\right)+5 (3 \delta m_N-\delta m_\Delta^1) \omega ^2 \left(\Sigma
-\omega ^2\right)\right) H^{\tilde \Lambda}(q)\nn &&+ 
\left.20 M_\pi^2 (3 \delta m_N-\delta m_\Delta^1) \Sigma
  L^{\tilde \Lambda} (q)\right) \,.
\eeqa
\end{itemize}
Here, $g_A$, $h_A$ and $\delta m_N^{\rm str}$ denote the nucleon, the
delta-nucleon axial-vector coupling and the strong contribution to
the neutron-proton mass splitting, in order.
The quantities $\Sigma$, $L^{\tilde \Lambda}$, $A^{\tilde \Lambda}$,
$D^{\tilde \Lambda}$ and  $H^{\tilde \Lambda}$ in the above expressions are
defined as follows: 
\beqa
\label{definitions}
\Sigma &=&  2M_\pi^2 + q^2 -2 \Delta^2\,, \nn
L^{\tilde \Lambda} (q) &=& \theta (\tilde \Lambda - 2 M_\pi ) \, \frac{\omega}{2 q} \, 
\ln \frac{\tilde \Lambda^2 \omega^2 + q^2 s^2 + 2 \tilde \Lambda q 
\omega s}{4 M_\pi^2 ( \tilde \Lambda^2 + q^2)}~, \quad
\omega = \sqrt{ q^2 + 4 M_\pi^2}~,  \quad 
s = \sqrt{\tilde \Lambda^2 - 4 M_\pi^2}\,, \nn
A^{\tilde \Lambda} (q) &=& \theta ( \tilde \Lambda - 2 M_\pi ) \frac{1}{2 q} 
\arctan \frac{q (\tilde \Lambda - 2 M_\pi )}{q^2 + 2 \tilde \Lambda M_\pi}\,, \nn
D^{\tilde \Lambda}(q) &=& \frac{1}{\Delta} \, \int_{2M_\pi}^{\tilde \Lambda}
\frac{d\mu}{\mu^2+q^2} \, \arctan \frac{\sqrt{\mu^2-4M_\pi^2}}{2\Delta}\,, \nn
H^{\tilde \Lambda}(q) &=& \frac{2 \Sigma }{\omega^2-4 \Delta^2} \biggl[
  L^{\tilde \Lambda}(q) - L^{\tilde \Lambda} (2\sqrt{\Delta^2 -
  M_\pi^2}) \biggr] \,.
\eeqa
Notice that the spectral function cutoff $\tilde \Lambda$ can be set to
$\infty$ in order to obtain the expressions corresponding to dimensional
regularization. We further emphasize that the factors of $\Sigma^{-1}$ 
in Eqs.~(\ref{triangle}-\ref{box_double}) always show up in the combination
$\Sigma^{-1} H^{\tilde \Lambda}(q)$ and thus do not lead to singularities.  

It is instructive to verify the consistency between the results 
obtained in EFT with and without explicit $\Delta$ degrees of freedom as done in
Ref.~\cite{Krebs:2007rh} for the isospin-conserving TPEP. 
Since both formulations differ from each other only by the different
counting of the delta-to-nucleon mass splitting, $\Delta \sim M_\pi \ll \Lambda_\chi$
versus $M_\pi \ll \Delta \sim \Lambda_\chi$ for the delta-full and the
delta-less theory, respectively. Thus expanding the various terms in
Eqs.~(\ref{triangle}-\ref{box_double}) in powers of $1/\Delta $  and counting $\Delta
\sim \Lambda_\chi$ should yield either terms polynomial in momenta
(i.e.~contact interactions) or non-polynomial contributions
absorbable into a  redefinition of the LECs in the delta-less theory
(in harmony with the decoupling theorem).
Expanding Eqs.~(\ref{triangle}-\ref{box_double}) in powers of $1/\Delta $
and keeping the $1/\Delta$ terms yields the following non-polynomial contributions:   
\beqa
V_S^{\rm II} &=& - q^2 \,  V_T^{\rm II} = \frac{g_A^2 h_A^2 \delta M_\pi^2 }{72 \pi
  F_\pi^4 \Delta}  q^2 A^{\tilde \Lambda}(q ) + \ldots \,, \nn
V_S^{\rm III} &=& - q^2 \, V_T^{\rm III} = \frac{g_A^2 h_A^2 \delta m_N }{72 \pi^2
  F_\pi^4 \Delta}  q^2 L^{\tilde \Lambda}(q ) + \ldots \,, \nn
V_C^{\rm III} &=& - \frac{g_A^2 h_A^2 \left(128 M_\pi^4+112 M_\pi^2 q^2+23 q^4\right)
  \delta m_N L^{\tilde \Lambda}(q)}{216 F_\pi^4 \pi ^2 (4 M_\pi^2 + q^2)
 \Delta }-\frac{h_A^2 \left(8 M_\pi^2+5 q^2\right) (2
\delta m_N-\delta m^{\rm str}) L^{\tilde \Lambda}(q)}{432 F_\pi^4 \pi ^2 \Delta } + \ldots\,,
\eeqa
where the ellipses refer to higher-order terms.
These expressions agree with the subleading IV contributions to the TPEP in
the delta-less theory given in Ref.~\cite{Epelbaum:2005fd} if one uses the values for the
LECs $c_i$ resulting from $\Delta$ saturation:
\beq
c_1 = 0\,, \quad \quad 
c_2 = - c_3 = 2 c_4 = \frac{4 h_A^2}{9 \Delta} \,.
\eeq
Notice that there is a factor $1/2$ missing in Eq.~(3.52) of
Ref.~\cite{Epelbaum:2005fd}. The correct expression for the central
component of the subleading CSB TPEP in the delta-less
theory $W_C^{(5)}$ reads 
\beqa
W_C^{(5)}&=&
-\frac{L^{\tilde \Lambda}(q)}{96\pi ^2F_\pi^4}
\left\{-g_A^2\delta m_N \frac{48M_\pi^4(2\,c_1+c_3)}{4 M_\pi^2 + q^2}\right.\nn &&+
4M_\pi^2\left[g_A^2\delta
  m_N(18\,c_1+2\,c_2-3\,c_3)+\frac{1}{2}(2\delta
  m_N-\delta m_N^{\rm str})(6\,c_1-
c_2-3\,c_3)\right]\nn && +
\left.q^2\left[g_A^2\delta m_N(5\,c_2-18\,c_3)-\frac{1}{2}(2\delta m_N-
\delta m_N^{\rm str})(c_2+6\,c_3)\right]\right\} ~.
\eeqa

%%%%%%%%%%%%%%%%%%%%%%%%%%%%%%%%%%%%%%%%%%%%%%%%%%%%%%%%%%%%%%%%%%%%%%%%%%%%%%%%%
\section{Results for the potential in configuration space}
\def\theequation{\arabic{section}.\arabic{equation}}
\setcounter{equation}{0}
\label{sec2}

We are now in the position to discuss the numerical strength of the obtained
IV TPEP and to compare the results with the ones arising in the delta-less
theory. The coordinate space representations of the various components of the
TPEP up to NNLO are defined according to   
\beqa
\tilde V (r) &=& \tau_1^3 \tau_2^3 \Big[ \tilde V_C^{\rm II} (r) + \tilde
V_S^{\rm II} (r)  \, 
\vec \sigma_1 \cdot \vec \sigma_2 + \tilde V_T^{\rm II}  (r) \, 
(3 \vec \sigma_1 \cdot \hat r \; \vec \sigma_2 \cdot \hat r  - 
\vec \sigma_1 \cdot \vec \sigma_2 ) \Big] + 
(\tau_1^3 + \tau_2^3) \Big[ \tilde V_C^{\rm III} (r) + \tilde
V_S^{\rm III} (r)  \, 
\vec \sigma_1 \cdot \vec \sigma_2 \nn
&& {} + \tilde V_T^{\rm III}  (r) \, 
(3 \vec \sigma_1 \cdot \hat r \; \vec \sigma_2 \cdot \hat r  - 
\vec \sigma_1 \cdot \vec \sigma_2 ) \Big]~.
\eeqa
The functions $\tilde V_{C,S,T}^{\rm II} (r)$ and $\tilde V_{C,S,T}^{\rm III}
(r)$ can be determined for any given $r > 0$ using the spectral function
representation as described in \cite{Kaiser:1998wa,Krebs:2007rh}. We use the
following values for the various LECs which appear in the TPEP:   
$g_A = 1.27$,  $h_A = 3g_A/(2\sqrt{2}) = 1.34$ from SU(4) (or large $N_c$),
$F_\pi = 92.4$~MeV,  $\delta M_\pi^2 = 1260$~MeV$^2$ and $\delta m_N =
-1.29$ MeV. For the strong nucleon mass shift, we adopt the value from 
Ref.~\cite{Gasser:1982ap}
$\delta m_N^{\rm str} = -2.05$ MeV,  see also \cite{Beane:2006fk} for a recent
determination from lattice QCD. The IV delta mass shifts $\delta m_\Delta^1$ and 
 $\delta m_\Delta^2$ have been determined in \cite{Epelbaum:2007sq} from the
physical values of
 $m_{\Delta^{++}}$, $m_{\Delta^{0}}$ and either the average delta mass 
$\bar{m}_{\Delta}=1233$~MeV leading to 
\beq
\label{delta_masses1}
\delta m_\Delta^1 = -5.3 \pm 2.0 \mbox{ MeV}, \quad \quad
\delta m_\Delta^2 = -1.7 \pm 2.7 \mbox{ MeV}
\eeq
or the quark mass relation  $m_{\Delta^+} -
m_{\Delta^0} = m_p - m_n$ leading to
\beq
\label{delta_masses2}
\delta m_\Delta^1 = -3.9 \mbox{ MeV} \, , \quad \quad
\delta m_\Delta^2 = 0.3 \pm 0.3 \mbox{ MeV}\,.
\eeq

\begin{figure}[tb]
  \begin{center} 
\includegraphics[width=16.6cm,keepaspectratio,angle=0,clip]{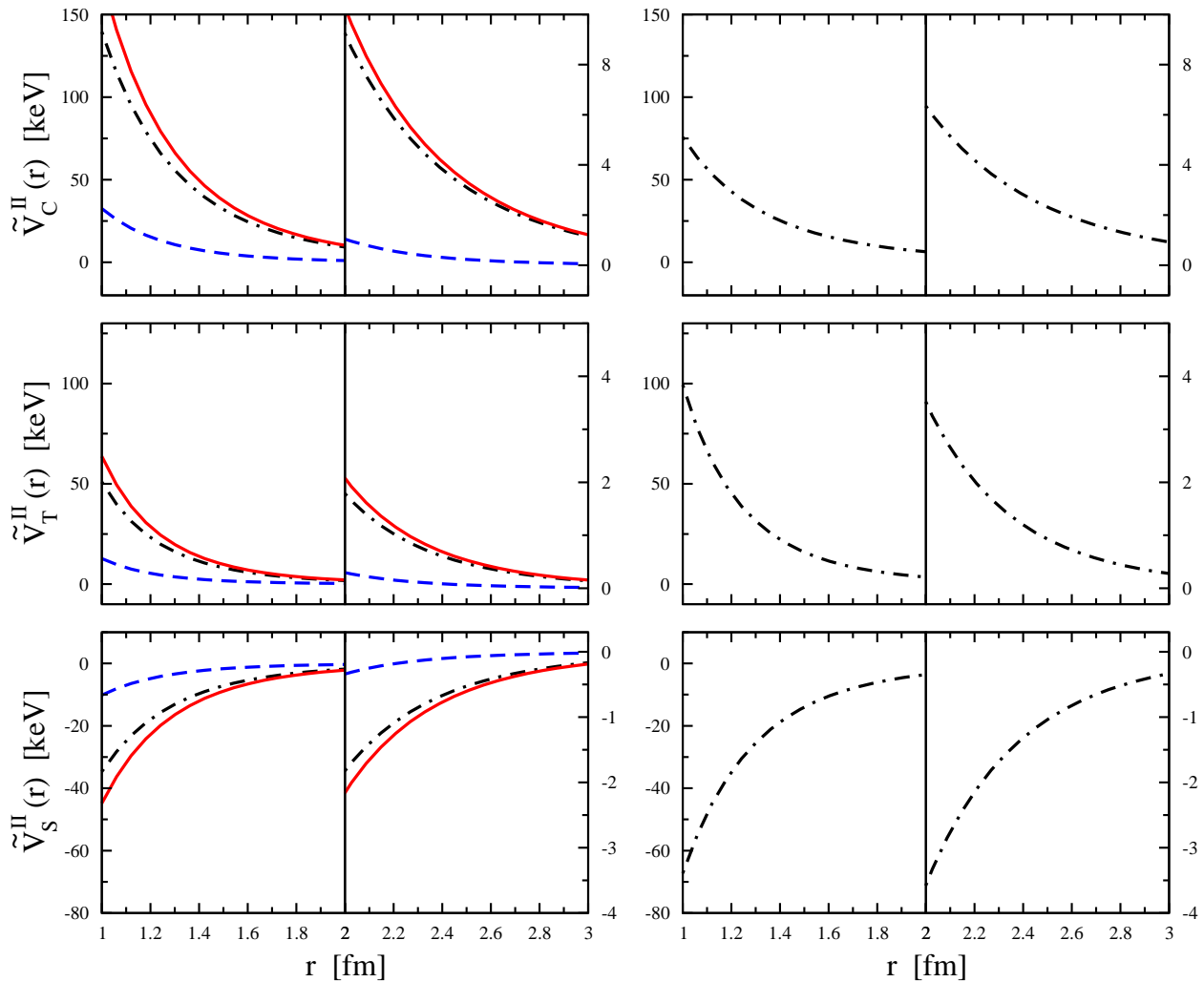}
    \caption{Class-II two-pion exchange potential. The left (right) panel shows the
      results obtained at leading order in chiral EFT with explicit
      $\Delta$ resonances (at subleading order in chiral EFT without explicit
      $\Delta$ degrees of freedom). The dashed and dashed-dotted lines depict the 
      contributions due to the delta and squared pion mass differences $\delta
      m_\Delta^2$ and $\delta M_\pi^2$, respectively, while the solid lines give
      the total result. In all cases, the spectral function cutoff $\tilde
      \Lambda = 700$~MeV is used. 
\label{v2} 
 }
  \end{center}
\end{figure}

Let us first discuss the charge-independence-breaking contributions to the
TPEP. In Fig.~\ref{v2}, 
we compare the strength of the corresponding central, spin-spin and tensor
components $\tilde V_{C,S,T}^{\rm II} (r)$ obtained at leading order in the
delta-full theory with the ones resulting at subleading order in the EFT 
without explicit delta.
In the former case, we add the leading-order contributions given in
Eq.~(3.40) of Ref.~\cite{Epelbaum:2005fd}, see also \cite{Friar:1999zr} for an
earlier calculation,  to the 
leading $\Delta$-contributions in Eq.~(\ref{triangle}-\ref{box_double}). 
In the latter case, we adopt the expressions given in
Ref.~\cite{Epelbaum:2005fd} and use the central values of the LECs $c_i$
found in Ref.~\cite{Krebs:2007rh}, namely:
\begin{equation}
c_1 = -0.57, \quad   
c_2 = 2.84, \quad  
c_3 = -3.87, \quad  
c_4 = 2.89,
\end{equation}
in units of GeV$^{-1}$. 
Notice that while in the delta-less theory, the leading and subleading class-II
TPEP arises entirely from the pion mass difference $\delta M_\pi^2$,  in the
delta-full theory one
also finds contributions proportional to $\delta m_\Delta^2$.
The results shown in Fig.~\ref{v2} for the contributions
$\propto \delta M_\pi^2$ are consistent with the observations made in
Ref.~\cite{Krebs:2007rh} for the isospin-invariant TPEP, namely that the next-to-leading
order (NLO) isovector central (spin-spin and tensor) components in the delta-full
theory are overestimated (underestimated) as compared to the
next-to-next-to-leading order (NNLO) calculation in the delta-less theory. We
remind the reader that the charge-independence-breaking TPEP due to the pion
mass difference can be expressed in terms of the corresponding
isospin-invariant TPEP as demonstrated in Ref.~\cite{Friar:1999zr}. 
The contributions proportional to $\propto \delta m_\Delta^2$ are numerically
smaller than the ones proportional to  $\delta M_\pi^2$ if one adopts the
central value $\delta m_\Delta^2 = -1.7$ MeV and lead to a slight
enhancement of the $\delta M_\pi^2$-contributions. Notice that the $\delta
m_\Delta^2$-terms provide a clear manifestation of effects which go beyond the
subleading order in the delta-less theory, see Ref.~\cite{Epelbaum:2007sq} for
a related discussion.   
\begin{figure}[tb]
  \begin{center} 
\includegraphics[width=16.6cm,keepaspectratio,angle=0,clip]{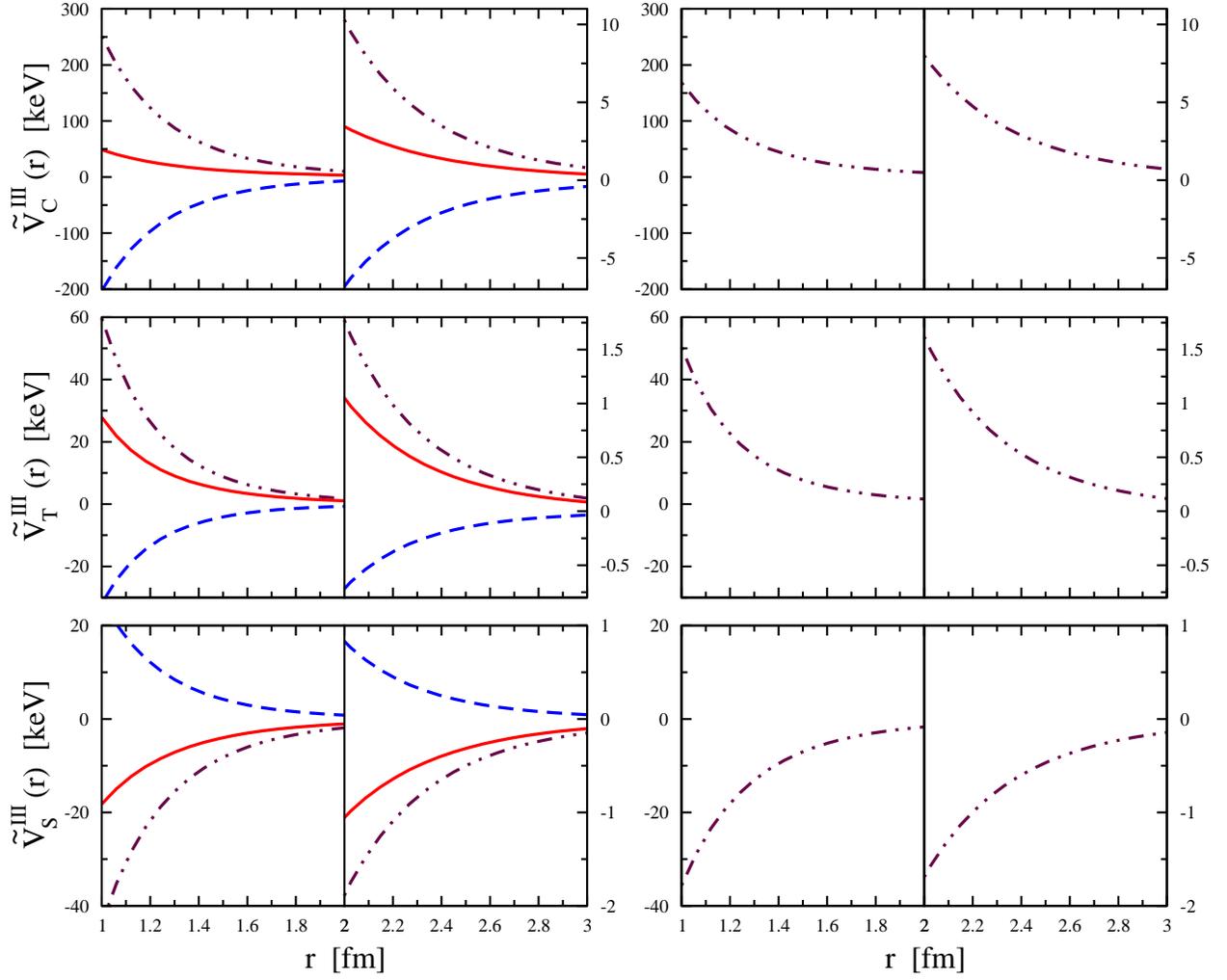}
    \caption{Class-III two-pion exchange potential. The left (right) panel shows the
      results obtained at leading order in chiral EFT with explicit
      $\Delta$ resonances (at subleading order in chiral EFT without explicit
      $\Delta$ degrees of freedom and assuming charge independence of 
      the $\pi N$ coupling constant, $\beta =0$). 
      The dashed and dashed-double-dotted lines depict the
      contributions due to the delta and nucleon mass differences $\delta
      m_\Delta^1$ and $\delta m_N^{\rm str, \; em}$, 
      respectively, while the solid lines give the total result. In all
      cases, the spectral function cutoff $\tilde 
      \Lambda = 700$~MeV is used. 
\label{v3} 
 }
  \end{center}
\end{figure}  

\begin{figure}[tb]
  \begin{center} 
\includegraphics[width=16.6cm,keepaspectratio,angle=0,clip]{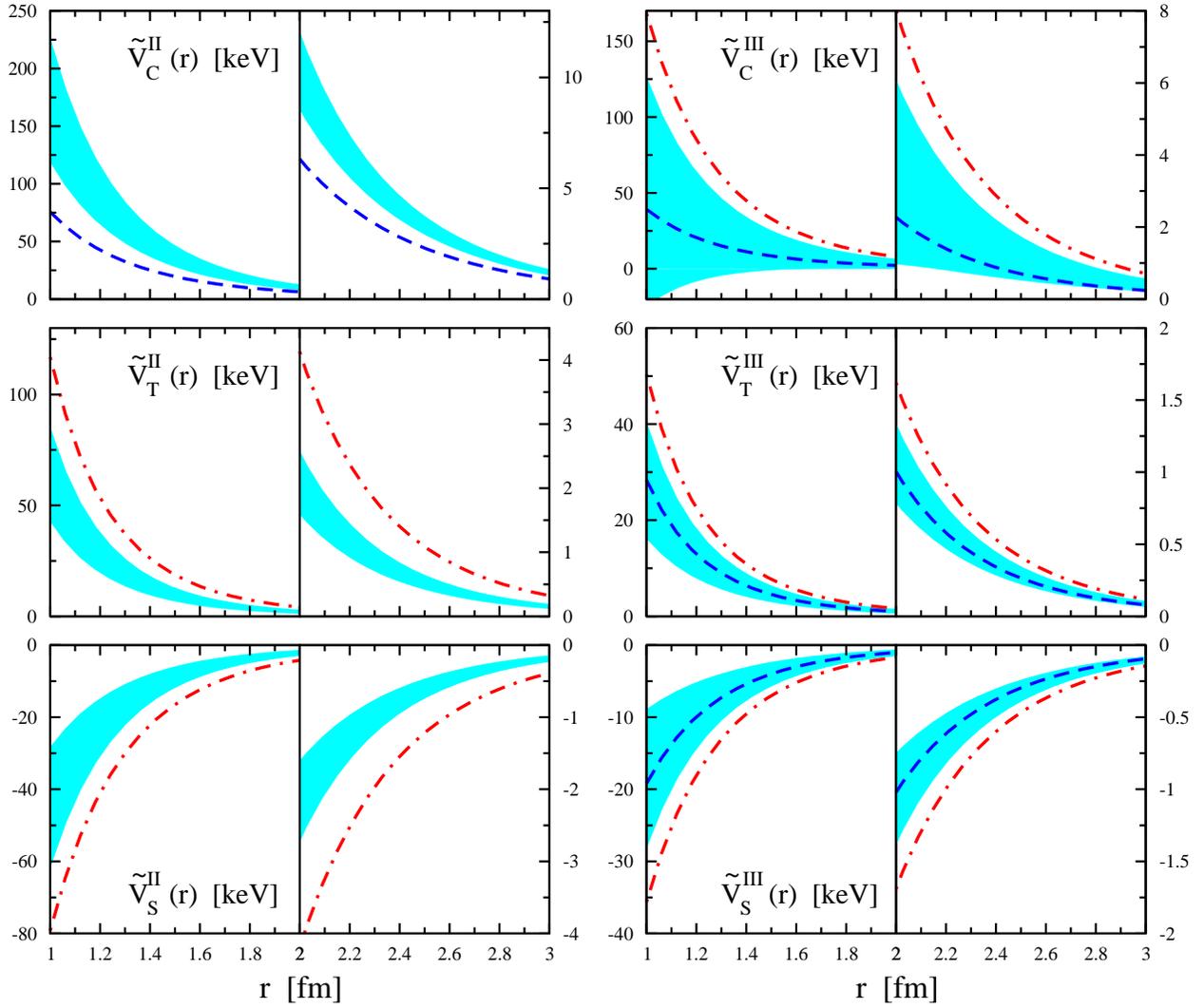}
    \caption{Class-II and class-III two-pion exchange
      potentials at leading order in the delta-full theory (shaded bands)
      compared to the results in the delta-less theory at leading (dashed
      lines) and subleading (dashed-dotted lines) orders. The bands arise from
      the variation of $\delta m_\Delta^1$ and $\delta m_\Delta^2$ according
      to Eq.~(\ref{delta_masses1}). Notice further that the leading
      (i.e.~order-$Q^4$) contributions to $\tilde V_{T,S}^{II}(r)$ and
      subleading (i.e.~order-$Q^5$) contributions to $\tilde V_{C}^{II} (r)$
      vanish in the delta-less theory. In all
      cases, the spectral function cutoff $\tilde 
      \Lambda = 700$~MeV is used. 
\label{v23} 
 }
  \end{center}
\end{figure}

Let us now regard the charge-symmetry-breaking TPEP. Again, we compare in
Fig.~\ref{v3} 
the leading-order results in the delta-full theory with the subleading
calculations in the EFT without explicit $\Delta$ using the results of
\cite{Epelbaum:2005fd}. The class-III TPEP is generated in the delta-less
theory by the strong and electromagnetic nucleon mass shifts and the charge
dependent pion-nucleon coupling constant $\beta$ \cite{Epelbaum:2005fd} whose
value is not known at present. In our numerical estimations, we set $\beta =
0$. In EFT with explicit $\Delta$ degrees of freedom, the class-II TPEP also
receives contributions proportional to the delta mass shift $\delta m_\Delta^1$.    
The $\delta m_N$-parts of $V_{S,T}^{\rm III}$ turn out to be very
similar in both cases while there are sizeable deviations for  $V_{C}^{\rm
  III}$. Notice that although the subleading contributions in the delta-full
theory have not yet been worked out and thus the convergence of the EFT expansion 
cannot yet be tested, the obtained results imply that the significant part of
the unnaturally big subleading contribution for the class-III TPEP in the
delta-less theory is now shifted to the lower order leading to a more
natural convergence pattern. The improved convergence of the delta-full theory
was also demonstrated for the isospin-invariant TPEP \cite{Krebs:2007rh}.  
In addition to the CSB terms generated by the nucleon mass shift, there are also
contributions proportional to $\delta m_\Delta^1$. For our central value,  
$\delta m_\Delta^1 = -5.3$ MeV, these contributions are numerically large and tend to
cancel the ones proportional to $\delta m_N$ and $\delta m_N^{\rm str}$
leading to a significantly weaker resulting class-III TPEP as compared to
the ones at subleading order in the delta-less theory. Similar cancellations were observed
recently for the IV 3NF \cite{Epelbaum:2007sq}.  This can be viewed as
an indication that certain higher-order IV contributions still missing at
subleading order in the delta-less theory are unnaturally
large in the theory without explicit delta degrees of freedom.  
We would further like to emphasize that there is a large uncertainty in the obtained
results for the IV TPEP due to the uncertainty in the values of  $\delta
m_\Delta^1$ and  $\delta m_\Delta^2$. This is visualized in Fig.~\ref{v23}
where the bands refer to the variation in the values $\delta
m_\Delta^{1,2}$ according to Eq.~(\ref{delta_masses1}).  

%%%%%%%%%%%%%%%%%%%%%%%%%%%%%%%%%%%%%%%%%%%%%%%%%%%%%%%%%%%%%%%%%%%%%%%%%%%%%%%%%
\section{Summary and conclusions}
\def\theequation{\arabic{section}.\arabic{equation}}
\setcounter{equation}{0}
\label{sec3}

In this paper we have studied the leading IV contributions to the TPEP  due to
explicit $\Delta$ degrees of freedom.  The pertinent results can be summarized
as follows: 
\begin{itemize}
\item[i)] We have calculated the triangle, box and crossed box NN diagrams
  with single and double delta excitations which give rise to the leading IV
  TPEP, see Fig.~\ref{fig1}. To facilitate the calculations, we used the 
  formulation based on the
  effective Lagrangian with the neutron-proton mass difference being
  eliminated. 
\item[ii)] We have verified the consistency of our results with the previous
  calculations based on the delta-less theory by expanding the non-polynomial
  contributions in powers of $1/\Delta$ and using resonance saturation for the
  LECs $c_i$.
\item[iii)] We found important contributions to the IV TPEP due to the mass
  splittings within the 
  delta quartet which go beyond the subleading order of the delta-less
  theory. In particular, the strong CSB potential found in
  Ref.~\cite{Epelbaum:2005fd} is significantly reduced by the contributions
  proportional to $\delta m_\Delta^1$. 
\end{itemize}
In the future, it would be interesting to derive  the subleading
$\Delta$-contributions to the IV TPEP in order to test the convergence of the
chiral expansion in the delta-full theory. The explicit expressions for the IV
TPEP worked out in this paper can (and should be) incorporated in the future
partial wave analysis of nucleon-nucleon scattering.

%%%%%%%%%%%%%%%%%%%%%%%%%%%%%%%%%%%%%%%%%%%%%%%%%%%%%%%%%%%%%%%%%%%%%%%%%%%%%%%%%
\section*{Acknowledgments}

The work of E.E. and H.K. was supported in parts by funds provided from the 
Helmholtz Association to the young investigator group  
``Few-Nucleon Systems in Chiral Effective Field Theory'' (grant  VH-NG-222)
and of U.M. through the virtual institute ``Spin and strong QCD'' (grant VH-VI-231). 
This work was further supported by the DFG (SFB/TR 16 ``Subnuclear Structure
of Matter'') and by the EU Integrated Infrastructure Initiative Hadron
Physics Project under contract number RII3-CT-2004-506078.

\setlength{\bibsep}{0.2em}
%\bibliographystyle{h-physrev3}
%\bibliography{/home/epelbaum/refs_h-elsevier3}

\end{document}